\documentclass[fleqn,usenatbib]{mnras}

\usepackage{newtxtext,newtxmath}
\usepackage{comment}

\usepackage[T1]{fontenc}

\usepackage{siunitx}
\DeclareRobustCommand{\VAN}[3]{#2}
\let\VANthebibliography\thebibliography
\def\thebibliography{\DeclareRobustCommand{\VAN}[3]{##3}\VANthebibliography}

\usepackage{graphicx}	
\usepackage{amsmath}	

\title[Prendergast Stability]{The Stability of Prendergast Magnetic Fields}

\author[Kaufman et al.]{
Emma Kaufman,$^{1}$\thanks{E-mail: ekaufman@u.northwestern.edu}
Daniel Lecoanet,$^{1,2}$
Evan H. Anders,$^{1}$
Benjamin P. Brown,$^{3}$
Geoffrey M. Vasil,$^{4}$
\newauthor
Jeffrey S. Oishi,$^{5}$
and Keaton J. Burns$^{6}$
\\
$^{1}$CIERA, Northwestern University, Evanston IL 60201, USA\\
$^{2}$Department of Engineering Sciences and Applied Mathematics, Northwestern University, Evanston IL 60208, USA\\
$^{3}$University of Colorado Department of Astrophysical and Planetary Sciences, Boulder, CO 80309, USA\\
$^{4}$School of Mathematics, University of Edinburgh, EH9 3FD, UK\\
$^{5}$Department of Physics \& Astronomy, Bates College, Lewiston, ME 04240, USA\\
$^{6}$Massachusetts Institute of Technology Department of Physics, Cambridge, MA 02139, USA
}

\date{Accepted XXX. Received YYY; in original form ZZZ}

\pubyear{2022}

\renewcommand{\vec}[1]{\ensuremath{\boldsymbol{#1}}}

\usepackage{xcolor}
\usepackage{ulem}

\makeatletter
\def\input@path{{./sections/}}
\makeatother
\graphicspath{{./}{figures/}}

\begin{document}
\label{firstpage}
\pagerange{\pageref{firstpage}--\pageref{lastpage}}
\maketitle

\begin{abstract}
Convection in massive main sequence stars generates large scale magnetic fields in their cores which persists as they evolve up the red giant branch. The remnants of these fields may take the form of the Prendergast magnetic field, a combination of poloidal and toroidal field components which are expected to stabilize each other. Previous analytic and numerical calculations
did not find any evidence for instability of the Prendergast field over short timescales. In this paper, we present numerical simulations which show a long timescale, linear instability of this magnetic field.
We find the instability to be robust to changes in boundary conditions and it is not stabilized by strong stable stratification. 
The instability is a \textit{resistive instability}, and the growth rate has a power-law dependence on the resistivity, in which the growth rate decreases as the resistivity decreases. 
We estimate the growth rate of the instability in stars by extrapolating this power-law to stellar values of the resistivity.
The instability is sufficiently rapid to destabilize the magnetic field on timescales shorter than the stellar evolution timescale, indicating that the Prendergast field is not a good model to use in studies of magnetic fields in stars.
\end{abstract}




\section{Introduction}

Stars have magnetic fields which influence dynamical processes relevant to stellar evolution such as stellar winds \citep[][]{stellarwinds}, chemical mixing \citep[][]{chmicakmixing}, heat transport \citep[][]{heattransport}, convection \citep[][]{convection}, and stellar pulsations \citep[][]{astroseismology}. 
Surface magnetic fields have been observed for stars in various stages of evolution \citep[][]{obs_bfield}. 
 
The presence of surface magnetic fields could be explained by two different hypotheses: the dynamo hypothesis and the fossil hypothesis. The dynamo hypothesis proposes that magnetic fields are generated within stellar convection zones by convective fluid motions \citep[][]{dynamofields1}. 
 The dynamo hypothesis explains surface observations in main-sequence solar type stars \citep[][]{solardynamo}, which have convective envelopes. 
 However, the dynamo hypothesis does not explain observations of surface magnetism in massive main-sequence stars \citep[][]{massive_no_dynamo}.
 These stars have convective cores and there is no established mechanism to transport the dynamo field across through the radiative envelope to the surface. 
 The fossil hypothesis proposes that stellar magnetic fields are remnants from earlier stages of stellar evolution \citep[][]{fossilfield1}. This hypothesis explains magnetism in massive stars \citep[][]{massive_no_dynamo}, although it requires a stable field configuration which could survive over the lifetime of the star. 

 The fossil hypothesis can also explain core magnetic fields in post main sequence stars such as red giant branch (RGB) stars. Astroseismic observations indicate strong core magnetic fields in RGB stars \citep[][]{astroseismology, stellocantiello}. These stars have a stably stratified, radiative core which was convective when the star was on the main sequence. 
 This convective history provides a dynamo mechanism which could have generated magnetic field on the main sequence. 
 In order for these fields to be observable on the RGB, the field configuration must be stable.
 This makes RGB stars a good model for investigating stable magnetic field configurations. 

 \begin{figure}
    \centering
    \includegraphics[width=\columnwidth]{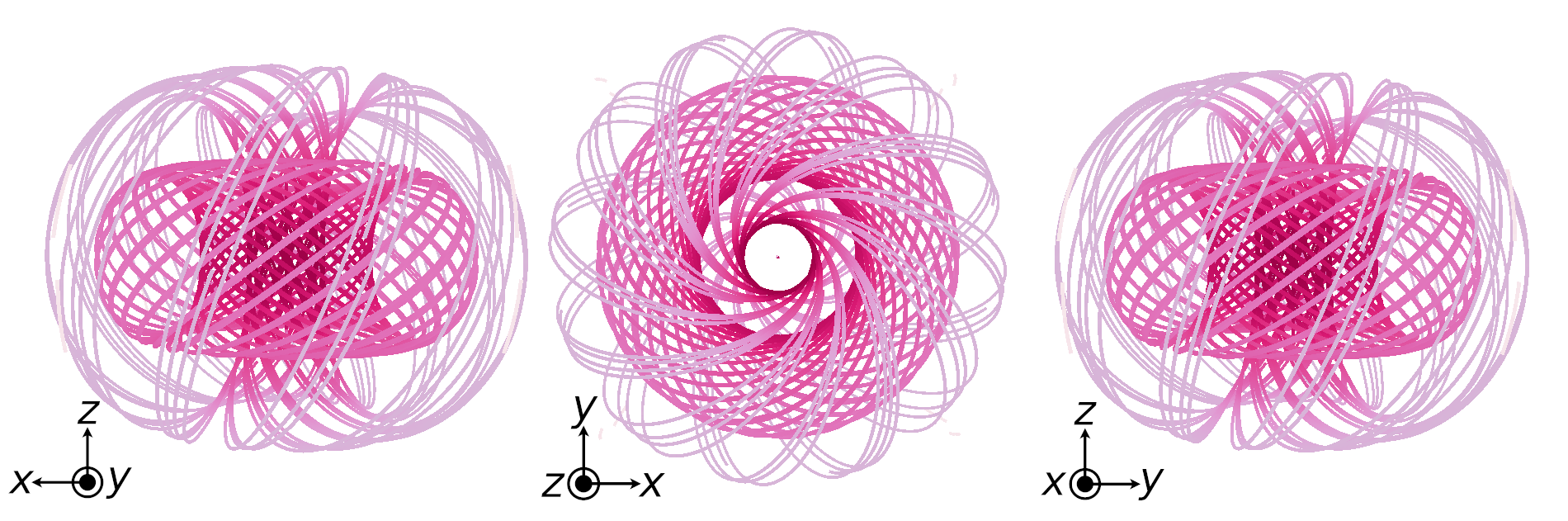}
    \caption{Magnetic field line visualization of the Prendergast magnetic field as given by equations \ref{eq:pren}-\ref{eq:prendend-sin/cos}. The color of the field lines indicates the strength of the magnitude of the field, with darker color indicating a stronger magnitude. The field consists of a combination of twisting toroidal and poloidal components with its axis of symmetry aligned with the z-axis. To aid in visualization we select 2 helical structures, but the field is continuous and has many nesting structures.}
    \label{fig:B0}
\end{figure}

 While purely toroidal \citep[][]{toroidal_stablility} and purely poloidal \citep[][]{poloidalstability} fields are unstable, a combination of the two may be stable. \citet{Prendergast1956} developed an equilibrium state with an axisymmetric mix of poloidal and toroidal magnetic field that we will call a ``Prendergast field" configuration. Figure \ref{fig:B0} shows a 3D visualization of the Prendergast field. The field consists of a combination of twisting toroidal and poloidal components and is axisymmetric about the z-axis. The magnitude of the field strength decreases radially. 
 
 Previous works have investigated the stability of the Prendergast field.
 \citet[][]{Prendergast2} showed via the variational principle that the $m=0$ mode is unstable once the magnetic energy exceeds 0.4 times the gravitational binding energy of the star. He notes that this is likely an overestimate of the allowed magnetic energy for a star. This theoretical criterion is shown to be sufficient for instability but not a necessary criterion. Therefore, this result gives no indication of the stability of the model below this limit. 
 Higher order modes were not studied because of the difficulty of the calculations.
 \citet[][]{cowling1960} used a similar approach, but focused on displacements near the outer boundary in a neutrally stratified star. He found the instability condition was met for all strengths of magnetic field, and concluded that the Prendergast field is always unstable in neutrally stratified stars. 
 
 More recently, \citet[][]{Duez2010} followed the nonlinear evolution of the Prendergast field over 10 Alfven times. They did not see any evidence of instability in this time frame.  
 \citet[][]{Braithwaite+nordlund2005} found for a stably stratified star that a random initial magnetic field configuration tends to decay into a field with mixed toroidal and poloidal components. They were able to follow the evolution for a few hundred Alfven times and found the magnetic energy to be roughly constant after the initial decay. Assuming this mixed toroidal and poloidal field was a Prendergast field, the authors then concluded that the Prendergast field is dynamically stable.
 In contrast, \citet[][]{mitchell2014} looked at random initial magnetic field configurations for neutrally stratified stars, and followed the evolution for a hundred Alfven times. They found that the neutrally stratified star never reached a stable equilibrium state in this time frame.
 This is evidence for an instability of the Prendergast field, consistent with the analytic work of \citet[][]{cowling1960}.

 The Prendergast magnetic field is widely used as a stable field configuration.  
 Prendergast fields have been used to investigate the effects of a core magnetic field on stellar oscillations in red giant stars \citep[][]{loi2017}, the effects of strong magnetic fields on the propagation of gravity waves in stellar interiors \citep[][]{loi2018}, the effects of magnetic fields on the frequency of gravity modes in rotating stars \citep[][]{prat2019}, and the effects of core magnetic field on the observable mixed-mode frequencies of stars \citep[][]{bugnet2021}.
 However, if the Prendergast field is unstable, then it may not be a good model for stellar magnetic fields.
 It is unclear if these previous results would still hold for a different magnetic field configuration.
 
Here we present a linear stability analysis of the Prendergast magnetic field configuration. We find that the Prendergast field exhibits linear instability, with kinetic energy growing exponentially regardless of the value of the diffusivity, the strength of stable stratification, or the boundary conditions. 

In Section \ref{numdeets}, we describe the setup of our simulations. Section \ref{instability} presents the underlying instability. 
Section \ref{diff} discusses its dependence on the diffusivity, section \ref{stablestrat} discusses the addition of stable stratification into the system, and section \ref{mandBC} discusses the effect of changing azimuthal wavenumber and boundary conditions. Section \ref{r0} discusses possible mechanisms for the instability. In section \ref{irlstars} we discuss implications of this instability for red giant stars.

\section{Numerical Details}\label{numdeets}
We want to determine if the Prendergast magnetic field is a good candidate for a fossil magnetic field configuration in the core of red giant stars. To do so, we examine a domain consisting of the radiative core and solve for linear perturbations about a background Prendergast magnetic field.  
We time-evolve a series of spherical simulations solving the magnetohydrodynamics equations in the Boussinesq approximation linearized about a background magnetic field $\vec{B}_0$, which has an associated current $\vec{J}_0$. We nondimensionalize lengthscales using the radius of the simulation region, $R$, which is the distance from the origin to the edge of the radiative zone. 
We nondimensionalize the magnetic field with the maximum of $|\vec{B}_0|$. This choice sets the nondimensional timescale to be the Alfven time, $t_A$, associated with this maximum magnetic field strength, where $t_A=\frac{R}{v_A}$, $v_A= \frac{\vec{B}_0}{\sqrt{4\pi \rho_c}}$, and $\rho_c$ is a constant reference density.
 We solve the magnetohydrodynamic equations under the Coulomb gauge, given by
\begin{align}
   \qquad 
    &\nabla \cdot \vec{u} = 0
    \label{eq:incompressibility_density}
    \\
    \begin{split}
    &\partial_t \vec{u} + \frac{1}{\rho_c}\nabla P - \vec{g}\frac{\rho'}{\rho_c} -\nu \nabla^2 \vec{u} = - \frac{1}{4 \pi \rho_c}\nabla^2 \vec{A} \times \vec{B}_0  \\
    & \quad \quad \quad \quad \quad \quad \quad \quad \quad \quad \quad \quad \quad \quad + \frac{1}{4 \pi \rho_c}\vec{J}_0 \times (\nabla \times \vec{A})
    \label{eq:momentum_density}
    \end{split}
    \\
    &\partial_t \vec{A} - \nabla \Phi - \eta \nabla^2 \vec{A} = \vec{u} \times \vec{B}_0
    \label{eq:mag_density}
    \\
    &\partial_t \rho' + \vec{u} \cdot \nabla \rho_0  - \kappa \nabla^2 \rho' = 0
    \label{eq:density}   
    \\
    &\nabla \cdot \vec{A} = 0
    \label{eq:coloumb gauge}
\end{align}
where $\vec{u}$ is the velocity, $P$ is the pressure, $\nu$ is the viscosity, $\vec{A}$ is the vector potential, $\Phi$ is the electric potential, $\eta$ is the magnetic resistivity, $\vec{g} = -gr\vec{e}_r$ is gravitational acceleration, $\rho ' $ is density perturbation, $\kappa$ is thermal diffusivity, and $\rho_0$ is the background density, which we take to vary as $\nabla \rho_0 = -2\rho_c r/R^2 \vec{e}_r$. In the Boussinesq approximation, the Brunt-V{\"a}is{\"a}l{\"a} frequency, $N^2$ is 
\begin{equation}
    N^2 = (-\partial_r \rho_0) \frac{g}{\rho_c}
    \label{eq:n2}
\end{equation}
In all cases $N^2$ is proportional to $r^2$, so we define $N_0^2$, the stable stratification strength, such that $N^2 = N_0^2 r^2$. 
In order to study the case where $N_0^2 = 0$, we set $\partial_r\rho_0=0$ and $\rho'=0$. 

We set our background magnetic field to the Prendergast profile outlined in \citet[][]{Prendergast1956, Loi}, 
\begin{equation}
    \begin{split}
            \vec{B}_0 &= \left(B_r, B_\theta, B_\phi\right)  \\
            &= \left(\frac{2}{r^2}\Psi(r) \cos \theta, -\frac{1}{r}\Psi'(r)\sin\theta,-\frac{\lambda}{r}\Psi(r)\sin\theta\right).    
    \label{eq:pren}   
    \end{split}
\end{equation}
The full field is a non-singular solution to the balance,
\begin{equation}
    \lambda \vec{B}_{0} + \nabla \times \vec{B}_{0} = - \beta r \sin(\theta) \vec{e}_{\!\phi} = \beta \, ( y \vec{e}_{\!x} - x \vec{e}_{\!y}) \label{eq:B0-balance},
\end{equation}
with $\vec{B}_{0} = \vec{0}$ at the outer boundary, $r=R=1$, and $\beta$ is the field amplitude.
In terms of the radial profile, 
\begin{equation}
    \Psi(r) = \frac{\beta}{\lambda^{2}}
    \left(r^{2} - r \frac{j_{1}(\lambda r)}{j_{1}(\lambda)}\right),
    \label{eq:prendend-sin/cos}
\end{equation} 
where $j_{1}(\xi) = \sin(\xi)/\xi^{2} - \cos(\xi)/\xi$ is a spherical Bessel function.
Automatically, $\Psi(r=1) = 0$.
The whole field vanishes at the outer boundary provided $\Psi'(r=1)=0$, requiring $\tan(\lambda) = 3\lambda/(3-\lambda^{2})$.
We choose the lowest-energy solution $\lambda \approx 5.76346$.
We set the overall amplitude $\beta = \lambda^{2} j_{1}(\lambda) /(2j_{1}(\lambda)-2\lambda/3) \approx 1.31765$, in accordance with our normalization.

We do not include resistivity in the leading-order magnetic balance. Using the stellar parameters from  \S\ref{irlstars}, the resistive timescale is $R^2/\eta =$\num{2.95e13} years, longer than the age of the universe.

We solve equations \ref{eq:incompressibility_density}-\ref{eq:coloumb gauge} using the Dedalus pseudospectral code \citep[][]{dedalus}, in spherical coordinates. For simulations using potential boundary conditions we run on commit 8639c65 of the d3 branch. For simulations using perfectly conducting boundary conditions we run on commit 4348a2f of the d3 branch. 
All variables are represented by spin-weighted spherical harmonics in the angular directions and radially weighted Jacobi polynomials in the radial direction \citep[][]{spheres_a,spheres_b}. The system we are studying is axisymmetric, so we are able to resolve the behavior in the $\phi$ direction even with a small number of points. Axisymmetry also allows the different $m$ modes to be independent of each other and thus studied separately. We utilize a 2nd-order semi-implicit SBDF scheme timestepper \citep[][]{SBDF2}.

We use two different types of magnetic boundary conditions: potential-field (POT) and perfectly conducting (PC). Potential boundary conditions 
assume that the external magnetic field is current-free, and that magnetic field is continuous at the boundary of
the simulation. 
We specify these conditions by decomposing $\vec{A}$ into spherical harmonic functions $\vec{A}_\ell$ where $\ell$ is the spherical harmonic degree and represent these conditions with $\partial_r\vec{A}_\ell + (\ell+1)\vec{A}_\ell / r = 0$. 
Perfectly conducting boundary conditions 
assume that the electric potential is constant at the boundary of the simulation
and that there is no normal magnetic field at the boundary. We impose these conditions with $\vec{e}_{\theta} \cdot \vec{A} = \vec{e}_{\phi} \cdot \vec{A} = \Phi = 0$.
In all cases we take the boundary to be stress free and impenetrable. We impose these conditions with 
$\vec{e}_r \cdot \vec{u} = \vec{e}_\theta \cdot S \cdot \vec{e}_r = \vec{e}_\phi \cdot S \cdot \vec{e}_r = 0$
where $S = \frac{1}{2} (\nabla\vec{u} + (\nabla \vec{u})^T )$ is the rank-2 stress tensor. We also take the density perturbation to be zero at $R=1$. 

Instabilities can be driven by an initial noise perturbation, forced at a certain frequency, or driven by a prescribed flow profile. Here, we focus on an instability driven by a prescribed flow profile. 
To seed the instability, we specify a velocity profile with a specific azimuthal wavenumber $m$. We define $\vec{u}_0=\sin (\theta)^m \sin(m\phi) e^{\frac{-(r-r_0)}{\Delta r^2}} \vec{e}_r$, where $r_0=0.875$ and $\Delta r=0.02$. Unfortunately $\vec{u}_0$ violates the incompressibility constraint, so we use divergence cleaning to find an incompressible initial condition associated with $\vec{u}_0$. We solve the equation 
\begin{equation}
    \nabla \cdot (\vec{u}_0 + \nabla P) = 0
\end{equation}
for the scalar field $P$ with boundary condition $\partial_r P(r=R)=0$. Then $\vec{u}_i=\vec{u}_0+\nabla P$
satisfies $\nabla \cdot  \vec{u}_i=0$ and the radial component of $\vec{u}_i$ goes to zero at $r=R$. We then initialize our simulation's velocity with $\vec{u}_i$.

We visualize the magnetic field structures and velocity flows using the Visualization and Analysis Platform for Ocean, Atmosphere, and Solar Researchers (VAPOR) \citep[][]{VAPOR}. We interpolate spherical data onto a Cartesian coordinate grid to create all the visualizations, which show bi-directional flow visualizations of desired fields.
To create the magnetic field visualizations, an even grid of seeds is created with 6 points in the direction of the symmetry axis and 3 points in the other directions. This distribution if seeds is chosen to limit the amount of field lines such that the central magnetic field structure is clear. In actuality, the magnetic field lines are continuous throughout the simulated region and form many concentric structures.
Each seed is integrated for 500 steps to find the magnetic field line. The color of the field lines indicates the strength of the magnitude of the field, with the darkest pink indicating the strongest magnetic field.
To create the velocity flow visualizations, we follow a similar procedure, but use 5 grid points in all directions and integrate for 200 steps. The color of the flow lines indicate the strength of the flow magnitude with darkest green indicating the highest magnitude flow.

\section{Results}
Here we present a linear stability analysis of the Prendergast magnetic field configuration. 
Section \ref{instability} presents the underlying instability. Section \ref{diff} discusses the effects of magnetic resistivity on the instability. 
Section \ref{stablestrat} discusses the addition of stable stratification into the system.
Section \ref{mandBC} discusses the effect of changing azimuthal wavenumber and boundary conditions on the instability and section \ref{r0} discusses possible mechanisms for the instability. 
In section \ref{irlstars} we discuss implications of this instability for red giant stars. 

\subsection{Instability of the Prendergast Field}\label{instability}
\begin{figure}
    \centering
    \includegraphics[width=\columnwidth]{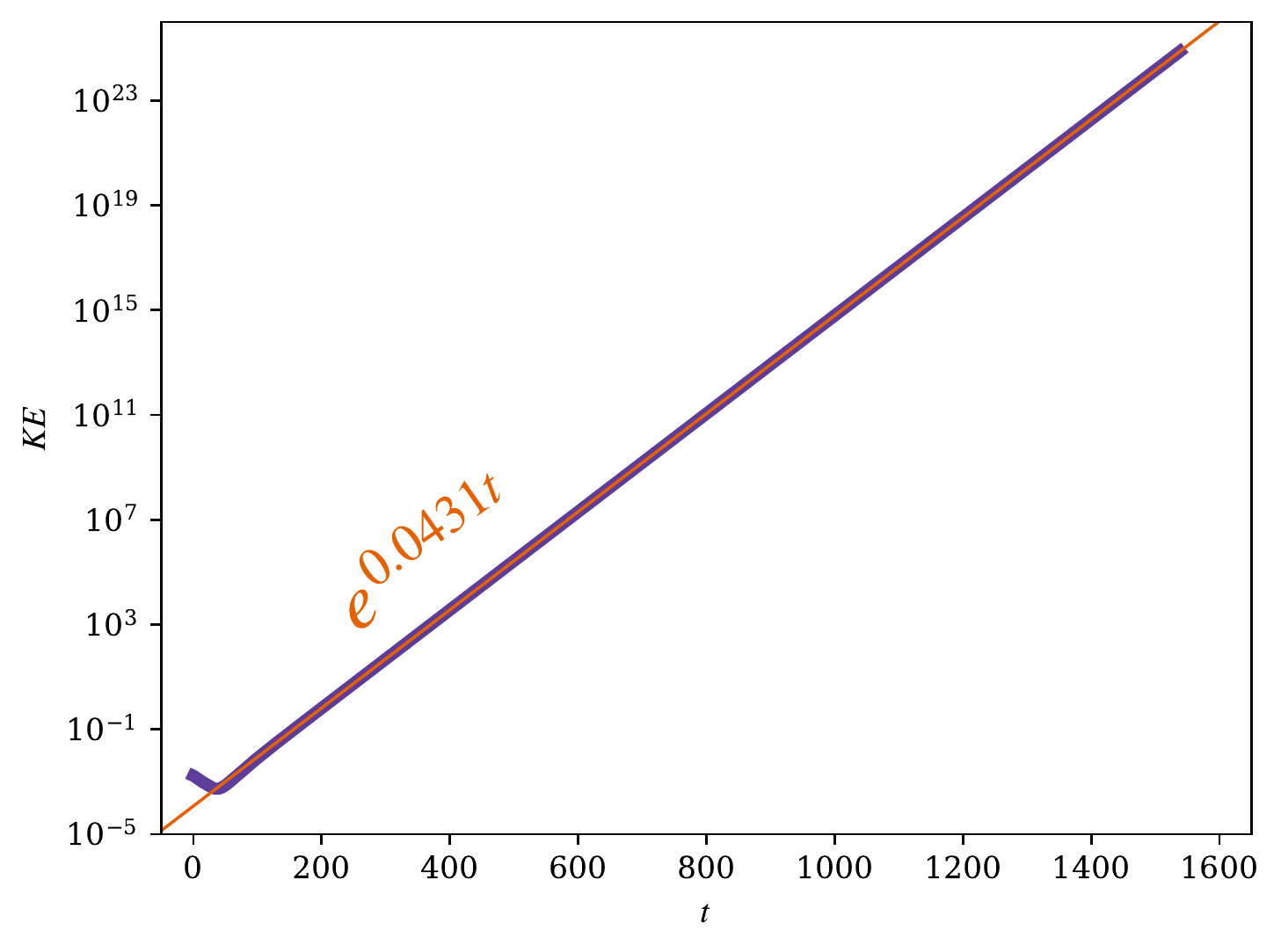}
    \caption{Kinetic energy as a function of time for the fiducial simulation. Simulation data is shown in purple and the exponential fit $KE = e^{0.0431t}$ is shown in orange.}
    \label{fig:ke}
\end{figure}

We time-evolve a series of three dimensional, spherical simulations of a radiative region with a background Prendergast magnetic field. Our fiducial case has a diffusivity of $\nu =$\num{1.3e-5}, no stable stratification ($N_0^2=0$), flow azimuthal wavenumber $m=1$, and potential boundary conditions. We run for 1600 Alfven times and calculate the total kinetic energy over time for the fiducial case. The total kinetic energy is given by $KE = 0.5 \left\langle u^2\right\rangle$, with angle brackets denoting a volume average. 
We plot kinetic energy as a function of time in figure \ref{fig:ke} with the simulation data shown in purple. We fit the exponential growth using a least squares linear fit given by $\ln(KE)=\gamma t$ where $\gamma$ is the growth rate.
The best fit is shown in orange. 
After the first 115 time units, once the instability has grown larger than the initial decaying state, kinetic energy grows steadily. 
We want to confirm that this instability is physical and not numerical. Our fiducial case has a growth rate of $\gamma=0.0431$, and we expect the growth rate to be indpendent of resolution and timestep size if the instability is physical. We first increase the spatial resolution to $N_r=N_{\theta}=255$, $N_{\phi}=4$ and find a growth rate of $\gamma=0.0433$ showing that the instability persists at higher resolution. We then decrease the timestep size to $10^{-4}$ and find a growth rate of $\gamma=0.0433$ showing that the instability persists at smaller timestep. Because the instability is robust even as we increase spatial resolution and decrease the timestep size (see table \ref{tab:sim_params_res}), we are confident that this instability is physical.  

\begin{figure*}
    \centering
    \includegraphics[width=0.9\textwidth]{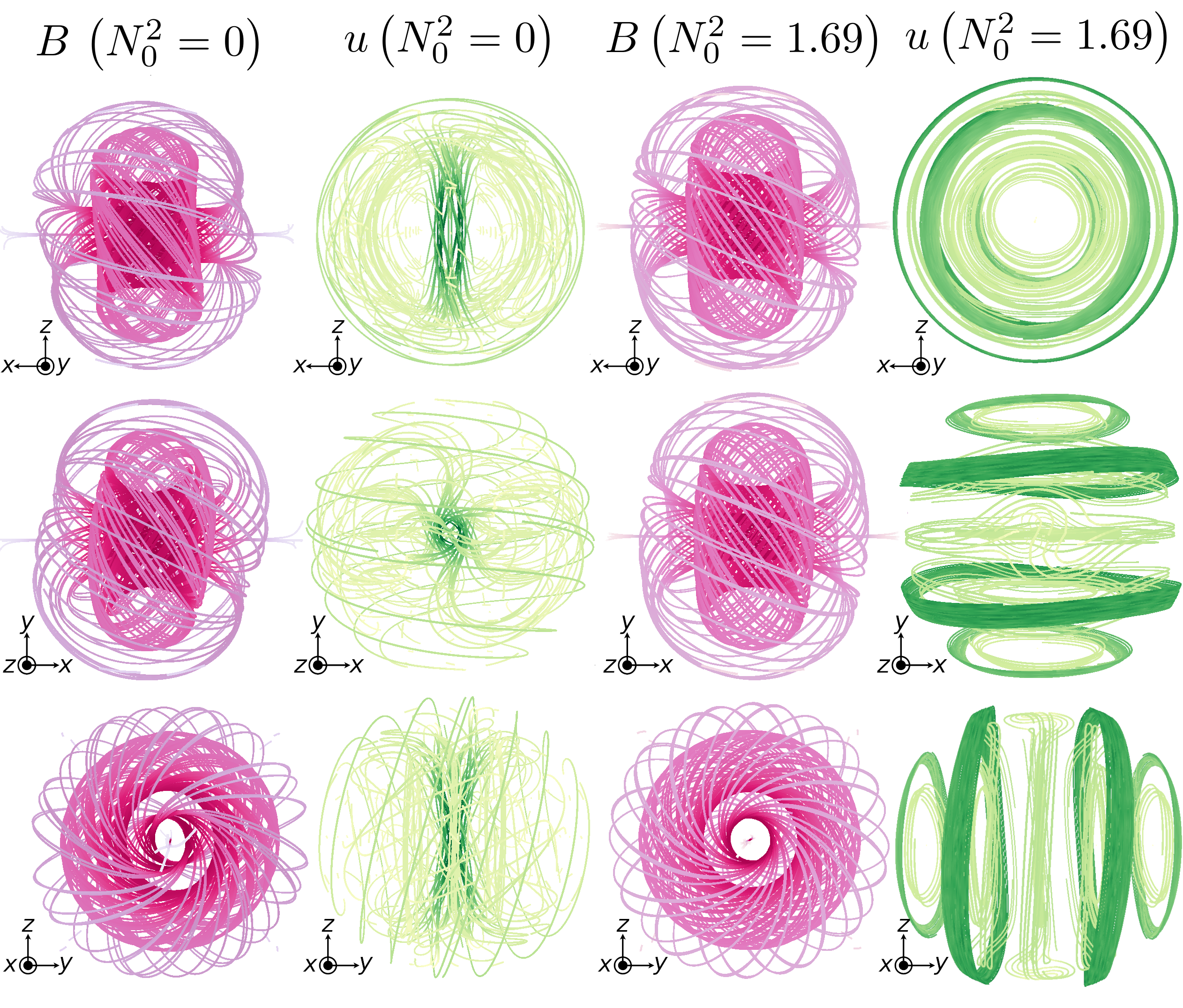}
    \caption{Columns (from left to right): Magnetic field eigenfunctions (pink)  for the $N_0^2=0$ case. Velocity flow eigenfunctions (green) for the $N_0^2=0$ case. Magnetic field eigenfunctions (pink) for the $N_0^2=1.69$ case. Velocity flow eigenfunctions (green) for the $N_0^2=1.69$ case.}
    \label{fig:eigs}
\end{figure*}

We visualize the magnetic field of the unstable eigenmode in the leftmost column of figure \ref{fig:eigs}.
We see a configuration very similar in shape to our background Prendergast field with a combination of twisting toroidal and poloidal components, but the axis of symmetry aligns with the x-axis, rather than the z-axis. While the background field is $m=0$ and thus must be symmetric about the z-axis, the perturbation has the same azimuthal wavenumber as the initial velocity flow, in this case $m=1$, and so cannot be symmetric around the z-axis, causing this flip in orientation for the unstable eigenmode. 
\begin{table}
	\centering
	\caption{Simulation parameters. Here $\eta$ is the magnetic resistivity, and $\eta = \nu = \kappa$. $N_r$, and $N_\phi$ are the r, and $\phi$ resolutions respectively. In all cases $N_r = N_\theta$. $\Delta t$ is the timestep size. $m$ is the azimuthal wavenumber. $N_0^2$ is the stable stratification strength. $\gamma$ is the growth rate of the instability. All simulations in this table utilize POT boundary conditions.}
	\label{tab:sim_params_res}
	\begin{tabular}{ccccccc} 
		\hline
		Name & $\eta$ & $(N_r, N_\phi)$ & $\Delta t$ & $m$ & $N_0^2$  & $\gamma$\\
		\hline
		Fid &\num{1.3e-5} & (127,4) & 0.0005 & 1 & 0  & 0.0431\\
		R255 &\num{1.3e-5} & (255,4) & 0.0005 & 1 & 0  & 0.0433\\
		dt &\num{1.3e-5} & (127,4) & 0.0001 & 1 & 0  & 0.0433\\
		\hline
	\end{tabular}
\end{table}
\subsection{Variation with Diffusivity}\label{diff}

\begin{figure}
    \centering
    \includegraphics[width=\columnwidth]{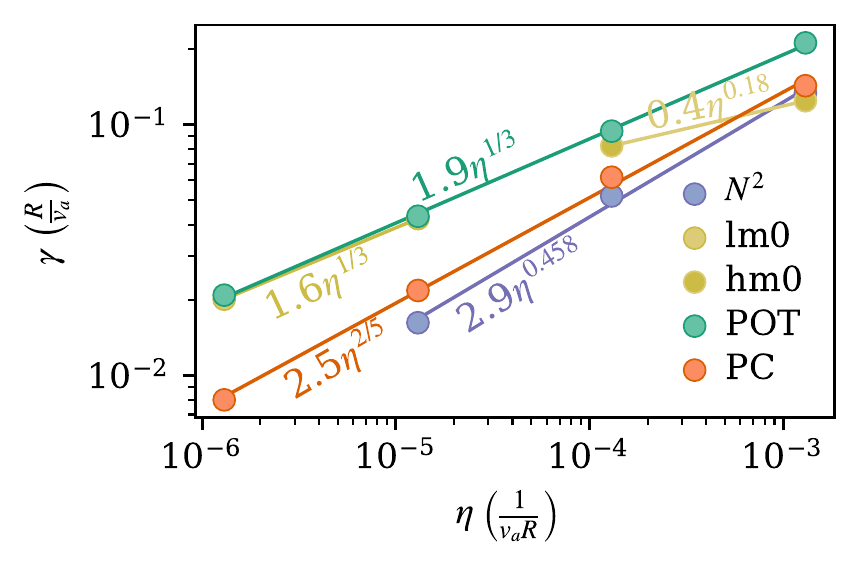}
    \caption{ Growth rate $\gamma$ as a function of resistivity $\eta$ for potential boundary conditions with $N_0^2=1.69$ and $m=1$ ($N^2$), potential boundary conditions with $N_0^2=0$ and $m=0$ at low (lm0) and high (hm0) resistivity, potential boundary conditions with $N_0^2=0$ and $m=1$ (POT), and perfectly conducting boundary conditions with $N_0^2=0$ and $m=1$ (PC). Each group of simulations is fit with a power law plotted as solid lines and given by the equation closest to each line. }
    \label{fig:nugamma}
\end{figure}

Our fiducial simulation is run at a resistivity much higher than we would expect in a physical star as it is infeasible to resolve a simulation with a realistic resistivity. However, we want to determine if this instability could still be present in physical systems by determining the effect of changing resistivity on the growth rate of the system. We ran 4 simulations with $\eta = \nu$ ranging from \num{1.3e-3} to \num{1.3e-6}. In table \ref{tab:sim_params_diff} we show the growth rates for each simulation and note that the growth slows as resistivity decreases, with growth rates ranging from 0.212 at the highest resistivity to 0.0209 at the lowest resistivity. The growth rate follows a power law as resistivity changes (see figure \ref{fig:nugamma}) with growth rate getting smaller as resistivity goes towards zero. 

Although resistivity often diffusively stabilizes perturbations, it can cause instability in certain systems. This is because the addition of resistive terms means plasma is no longer constrained to flow along the magnetic field lines. Resistive instabilities, such as the tearing mode instability \citep[][]{tearing}, draw on magnetic energy generated by plasma currents as the magnetic field tries to relax into a lower energy state. We discuss the possibility of a tearing mode like instability of the Prendergast magnetic field in section \ref{r0}. 

\begin{table}
	\centering
	\caption{See table \ref{tab:sim_params_res} for column descriptions. All simulations in this table use POT boundary conditions}
	\label{tab:sim_params_diff}
	\begin{tabular}{ccccccc} 
		\hline
		Name & $\eta$ & $(N_r, N_\phi)$ & $\Delta t$ & $m$ & $N_0^2$ & $\gamma$\\
		\hline
		D3 & \num{1.3e-3} & (127,4) & 0.0005 & 1 & 0  & 0.212\\
		D4 & \num{1.3e-4} & (127,4) & 0.0005 & 1 & 0  & 0.0943\\
		Fid &\num{1.3e-5}& (127,4) & 0.0005 & 1 & 0  & 0.0431\\
		D6 & \num{1.3e-6}& (511,4) & 0.0001 & 1 & 0  & 0.0209\\
		\hline
	\end{tabular}
\end{table}

\subsection{Introduction of Stable Stratification}\label{stablestrat}
Up until now our simulations have not included stable stratification, but red giant stars are stably stratified in the core region where we expect magnetic fields to live. 
Stable stratification damps out radial motions, and thus is expected to stabilize instabilities.
\begin{table}
	\centering
	\caption{See table \ref{tab:sim_params_res} for column descriptions. All simulations in this table use POT boundary conditions}
	\label{tab:sim_params_n2}
	\begin{tabular}{ccccccc} 
		\hline
		Name & $\eta$ & $(N_r, N_\phi)$ & $\Delta t$ & $m$ & $N_0^2$ & $\gamma$\\
		\hline
    	Fid &\num{1.3e-5}& (127,4) & 0.0005 & 1 & 0  &         0.0432\\
    	N0.00169& \num{1.3e-5}& (127,4) & 0.0005 & 1 & 0.00169 & 0.0420\\
    	N0.0169&\num{1.3e-5} & (127,4) & 0.0005 & 1 & 0.0169 &  0.0342\\
    	N0.0338&\num{1.3e-5} & (127,4) & 0.0005 & 1 & 0.0338 &  0.0277\\
    	N0.0845&\num{1.3e-5} & (127,4) & 0.0005 & 1 & 0.0845&  0.0166\\
    	N0.169&\num{1.3e-5} & (127,4) & 0.0005 & 1 & 0.169 &  0.0165\\
    	N0.338&\num{1.3e-5} & (127,4) & 0.0005 & 1 & 0.338  & 0.0164\\
    	N0.676&\num{1.3e-5} & (127,4) & 0.0005 & 1 & 0.676  & 0.0164\\
    	N1.014&\num{1.3e-5} & (127,4) & 0.0005 & 1 & 1.014  & 0.0164\\
        N1.69&\num{1.3e-5}& (127,4) & 0.0005 & 1 & 1.69  &      0.0163\\
        N16.9&\num{1.3e-5} & (127,4) & 0.0005 & 1 & 16.9 &    0.0163\\
        N169&\num{1.3e-5} & (127,4) & 0.0005 & 1 & 169 &     0.0163\\
        
		\hline
	\end{tabular}
\end{table}

In order to determine the effect of stable stratification strength on the growth rate of the kinetic energy, we change the simulation parameters such that $N_0^2$ varies from $0$ to $169$ (see table \ref{tab:sim_params_n2}). $N_0^2$ is a measure of the strength of stable stratification. The growth rates range from 0.0432 for $N_0^2=0$ to 0.0163 for $N_0^2=169$.
As described in section 3.1, we confirm that the instability is not altered by changing the resolution and timestep of the simulation. We note that the magnetic field of the unstable mode has roughly the same shape and orientation as seen in the case without stable stratification (see Figure \ref{fig:eigs}) indicating that the addition of stable stratification has no effect on the overall shape of the magnetic field of the unstable mode. 

\begin{figure}
    \centering
    \includegraphics[width=\columnwidth]{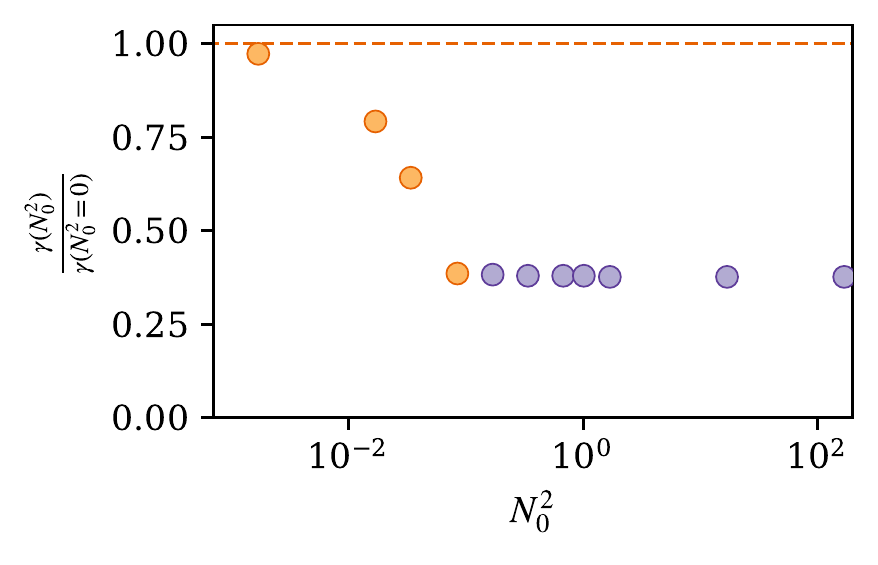}
    \caption{Growth rates in simulations with stable stratification normalized by the growth rate in the unstratified ($N_0^2 = 0$) case are plotted as a function of stable stratification strength, $N_0^2$. Orange dots indicate simulations at low $N_0^2$ where growth rate changes significantly as $N_0^2$ increases. Purple dots indicate simulations at high $N_0^2$ where growth rate is roughly constant as $N_0^2$ increases. The dashed orange line indicates unity.}
    \label{fig:n2_gamma}
\end{figure}

We plot the growth rate normalized to the unstratified case ($N_0^2=0$) as a function of $N_0^2$ in figure \ref{fig:n2_gamma}. 
We find that for low $N_0^2$, shown in orange, $\gamma$ changes as $N_0^2$ changes. For high $N_0^2$, shown in purple, $\gamma$ is roughly constant as $N_0^2$ changes. Utilizing a typical value of $N=10^3 \mu$Hz from \citet[][]{astroseismology} and the stellar parameters given in section \ref{irlstars}, we find that a typical value of $N_0^2$ for stars would be \num{3e11} in our simulations. While this is much higher than our simulated values, the constant growth rate at high $N_0^2$ indicates that the behavior at stellar values will be similar to the behavior at our highest simulated $N_0^2$.  
\begin{figure}
    \centering
    \includegraphics[width=\columnwidth]{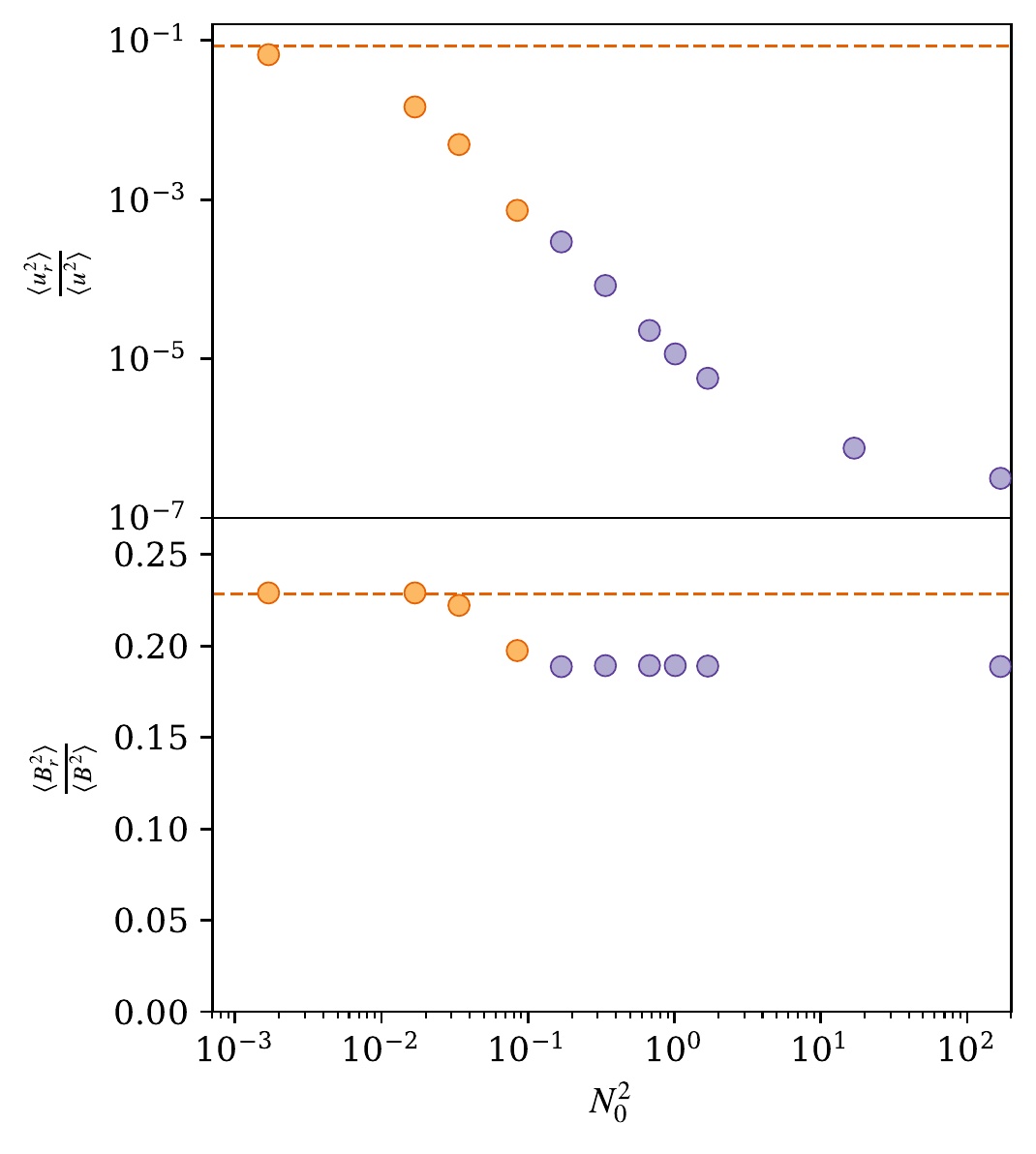}
    \caption{The top panel shows $\left \langle u_r^2 \right \rangle / \left \langle u^2\right \rangle$ as a function of $N_0^2$. The horizontal line shows the fraction of kinetic energy contained in the radial component for $N_0^2=0$. This fraction decreases over many orders of magnitude as $N_0^2$ increases. 
    The bottom panel shows $\left \langle B_r^2\right \rangle/ \left \langle B^2\right \rangle$ as a function of $N_0^2$. The horizontal line shows the fraction of magnetic energy contained in the radial component for $N_0^2=0$. This fraction stays fairly constant as $N_0^2$ increases.}
    \label{fig:N2_ur_br}
\end{figure}

Although we originally assumed that there would be some cutoff point where the stable stratification overpowered the magnetic instability, our results seem to indicate that the presence of a Prendergast magnetic field will cause an instability in a stably stratified star, regardless of the strength of the stratification relative to the strength of the magnetic field. There appear to be two separate regions of instability: the low $N_0^2$ region over which $\gamma$ changes with $N_0^2$ and the high $N_0^2$ region over which $\gamma$ is roughly constant. Figure \ref{fig:eigs} shows the magnetic field lines (pink) and velocity flow lines (green) for our low $N_0^2$ region (left) and high $N_0^2$ region (right). We find that the magnetic field lines are almost identical in the low and high $N_0^2$ regimes while the velocity flow lines are visibly different. The low $N_0^2$ regime has a radially dominated velocity flow, with fast helical flows through the central region. The high $N_0^2$ regime has almost no radial flows. This is seen in the eigenmodes which have no flow through the center of the star, only concentric circular flows. 

We quantify the difference in radial velocities in the top panel of figure \ref{fig:N2_ur_br} which shows the ratio of radial kinetic energy to total kinetic energy as a function of $N_0^2$. We see that this ratio decreases by 6 orders of magnitude as we go from low $N_0^2$ to high $N_0^2$. 
In contrast, the bottom panel of figure \ref{fig:N2_ur_br} shows the ratio of radial magnetic energy to total magnetic energy as a function of $N_0^2$. We see that this ratio changes by less than an order of magnitude over the same range of $N_0^2$.
The ratio $\langle B_r^2\rangle/\langle B^2 \rangle$ and the growth rate $\gamma$ have similar variation with $N_0^2$ (figure \ref{fig:n2_gamma}). This leads to a question; how is the radial magnetic field maintained in simulations with very small radial velocities? 

To answer this question we want to find an equation for the radial magnetic energy. We start with the radial component of equation \ref{eq:mag_density} and then multiply by the radial component of the magnetic field. We then take the volume average and find 
\begin{equation}
    \partial_t \left\langle B_r^2\right\rangle - \eta \nabla^2 \left\langle B_r^2\right\rangle = \left\langle \vec{e}_r \cdot \nabla \times \left( \vec{u} \times \vec{B}_0 \right) B_r\right\rangle
    \label{eq:magenergy}
\end{equation}
which, assuming axisymmetry, becomes
\begin{equation}
    \partial_t \left\langle B_r^2\right\rangle - \eta \nabla^2 \left\langle B_r^2\right\rangle = \left\langle  r^{-1}   \left( u_r B_{0,\theta} - u_\theta B_{0,r} \right) \partial_\theta B_r \right\rangle
\end{equation}
We see that the radial component of the magnetic energy depends on both $u_r$ and $u_\theta$. For high $N_0^2$, $u_r \rightarrow 0$, but the $u_\theta$ term does not approach zero as $N_0^2$ changes. This means that almost all of the radial magnetic energy in our high $N_0^2$ cases must come from the $\theta$ component of the velocity.
To quantify this, we take the right hand side of equation \ref{eq:magenergy} normalized with respect to magnetic energy. We find that this quantity changes from 0.008 at $N_0^2\sim 0$ to 0.002 at large $N_0^2$. This indicates that the radial magnetic energy is held roughly constant as we increase $N_0^2$, despite the difference in velocity flow lines with changing $N_0^2$.
Since the growth rate, radial magnetic energy, and right hand side of equation \ref{eq:magenergy} don't change significantly over a large range of $N_0^2$, we predict that this instability will persist across all strengths of stable stratification. While we are not able to simulate out the the high $N_0^2$ seen in stars, we predict that the behavior we see at the largest simulated value will be similar to what we see in stars because of all of these values become roughly constant across a large range of $N_0^2$. 

We know from \citet[][]{Prendergast2} that the Prendergast field will be unstable if the magnetic energy exceeds 0.4 times the gravitational binding energy of the star, although Prendergast notes that this is a physically irrelevant limit. In the Boussinesq approximation the gravitational binding energy is infinite and so the magnetic energy is effectively zero times the gravitational binding energy.
That we see an instability even with magnetic energy at zero times the gravitational binding energy indicates that Prendergast's upper limit overestimates the stability of the model, and that the Prendergast field will be unstable at all strengths of magnetic energy for both uniform and stably stratified stars. 

\subsection{Effect of Azimuthal Wavenumber and Boundary Conditions}\label{mandBC}
In our simulations, we are able to select for a particular azimuthal wave number $m$ of the initial velocity perturbations. However, in a star there are perturbations which take on a variety of $m$ values. Here we investigate the effect of $m$, the azimuthal wavenumber, on the growth rate of the kinetic energy. We select for different $m$ by initializing with $\vec{u}_0 = \sin (\theta)^m \sin(m\phi) e^{\frac{-(r-r_0)}{\Delta r^2}} \vec{e}_r$ and applying divergence cleaning to find an incompressible initial velocity. For these simulations, every 10,000 timesteps we zero the perturbations at all values of $m$ except for the one being studied. We do this because even growth of an unstable mode seeded at machine precision can be enough to overtake the kinetic energy of the system during the full time-integration of our simulation. We find that the $m=0$ and $m=1$ modes are unstable, with positive growth rate, while for all $m>1$, $\gamma$ is negative, i.e. the kinetic energy is decreasing over time (see table \ref{tab:sim_params_m}). This indicates that only the $m=1$ and $m=0$ modes are unstable. 
\begin{table}
	\centering
	\caption{See table \ref{tab:sim_params_res} for column descriptions. All simulations in this table use POT boundary conditions.}
	\label{tab:sim_params_m}
	\begin{tabular}{ccccccc} 
		\hline
		Name & $\eta$ & $(N_r, N_\phi)$ & $\Delta t$ & $m$ & $N_0^2$ & $\gamma$\\
		\hline
		m0 &\num{1.3e-5}& (127,4) & 0.0005 & 0 & 0  & 0.0421\\
    	Fid & \num{1.3e-5}& (127,4) & 0.0005 & 1 & 0  & 0.0412\\
    	m2 &\num{1.3e-5} & (127,8) & 0.0005 & 2 & 0 &  -0.000650\\
    	m3 &\num{1.3e-5} & (127,16) & 0.0005 & 3 & 0 &  -0.00915\\
		\hline
	\end{tabular}
\end{table}

\begin{figure}
    \centering
    \includegraphics[width=\columnwidth]{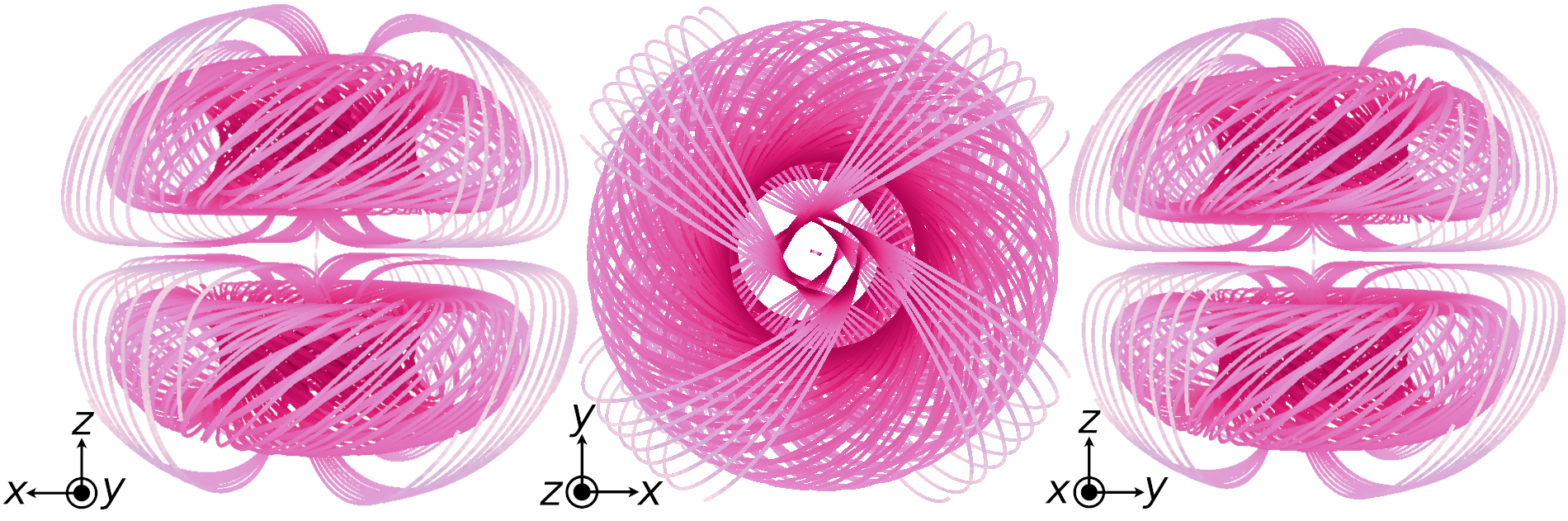}
    \caption{The unstable mode for azimuthal wavenumber $m=0$. Darker color indicates a stronger magnetic field magnitude.}
    \label{fig:m0_vis}
\end{figure}

We visualize the unstable $m=0$ mode in Figure \ref{fig:m0_vis} using VAPOR.  We see that the magnitude of the magnetic field is again increasing towards the center, but the unstable mode is split across the x-y plane, showing two distinct but identical halves. 
In figure \ref{fig:nugamma}, we plot the growth rate as a function of resistivity for the $m=0$ mode and find that the $m=0$ mode follows a broken power law. At low resistivity, the slopes are the same for the $m=1$ and $m=0$ modes, whereas the $m=1$ mode has a steeper slope than the $m=0$ mode at low resistivity. We discuss possible explanations for this difference in slope in section \ref{r0}. 

Next we determine the effects of our boundary condition choice on the growth rate of the instability. 
We find that the instability is present for both  potential and perfectly conducting boundary conditions, although the growth rate is greater for potential than perfectly conducting across all diffusivities (see figure \ref{fig:nugamma}). The power law index differs for the different boundary conditions.
It is known that boundary condition choice can have a significant effect on the growth rate of other magnetic instabilities, like magnetic dynamos \citep[][]{varela, Thelen}. \citet[][]{fontana} found that perfectly conducting boundary conditions had a larger growth rate than potential boundary conditions for magnetic dynamos in direct opposition to our findings here. 
However, a major difference between those previous studies and our own is that the Prendergast magnetic field is unstable to a resistive instability.

\begin{table}
	\centering
	\caption{See table \ref{tab:sim_params_res} for column descriptions. All simulations in this table use PC boundary conditions.}
	\label{tab:sim_params_PCBC}
	\begin{tabular}{ccccccc} 
		\hline
		Name & $\eta$ & $(N_r, N_\phi)$ & $\Delta t$ & $m$ & $N_0^2$ & $\gamma$\\
		\hline
		PC-D3 &\num{1.3e-3}& (128,4) & 0.0005 & 1 & 0  & 0.143\\
    	PC-D4 & \num{1.3e-4}& (128,4) & 0.0005 & 1 & 0  & 0.0617\\
    	PC-D5 &\num{1.3e-5} & (128,4) & 0.0005 & 1 & 0 &  0.0218\\
    	PC-D6 &\num{1.3e-6} & (512,4) & 0.0005 & 1 & 0 &  0.00801\\
		\hline
	\end{tabular}
\end{table}

\subsection{Interpretation of the Instability}\label{r0}
Here we discuss the tearing mode instability as a potential mechanism for the instability seen in the Prendergast magnetic field.
The tearing mode instability is a resistive instability that occurs at boundary layers, places where the magnetic field is zero \citep[][]{tearing}. 
Tearing mode instabilities can also occur when a guide field is present at boundary layers \citep[][]{drake1, drake2}. A guide field is a nonzero magnetic field normal to the boundary layer. 
The Prendergast magnetic field is zero at $r=0$ in all components except the z component. 
While boundary layers are typically two dimensional, the Prendergast field has a zero dimensional ``boundary layer'' with a guide field at $r=0$, since the magnetic field goes to zero in all but one component. 

The eigenmodes of the tearing mode instability have two distinct features. The magnetic field perturbations normal to the boundary layer are continuous across the boundary layer, while the  normal derivative of the normal field is discontinuous across the boundary layer in the limit that $\eta\rightarrow 0$.
If the instability of the Prendergast magnetic field also has sharp gradients in the normal magnetic field, that would suggest it is a tearing mode-like instability.

\begin{figure}
    \centering
    \includegraphics[width=\columnwidth]{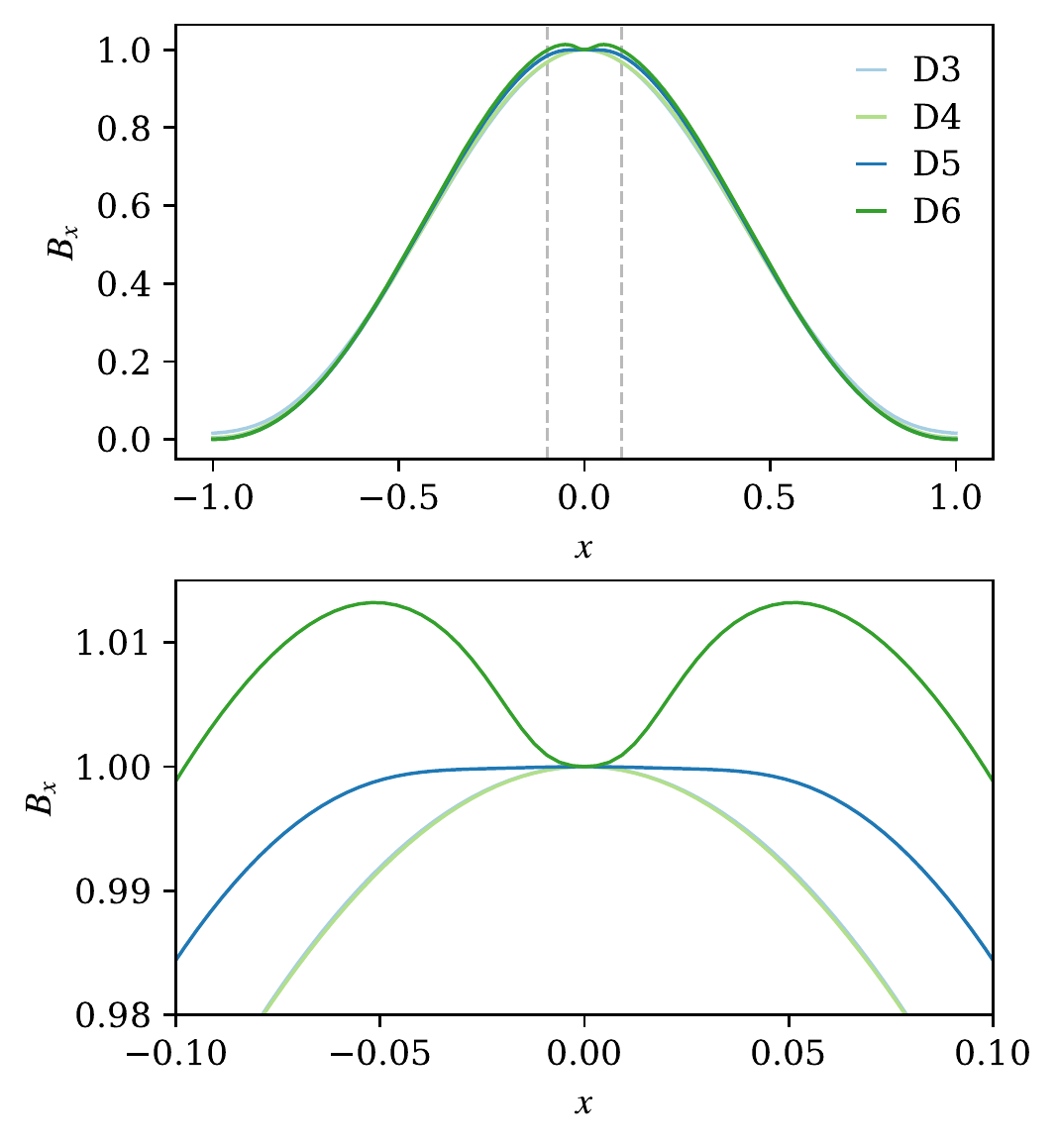}
    \caption{Top Panel: $B_x$ as a function of $x$ along the $x$-axis for $m=1$ simulations with varying resistivities, where D3 corresponds to a resistivity of $\num{1.3e-3}$ and so on. In all cases the simulations have potential boundary conditions. The vertical dashed lines denote the region of interest for the bottom panel. Bottom Panel: As above, but zoomed into the boundary layer around $r=0$.}
    \label{fig:diffbl}
\end{figure}

We see in figure \ref{fig:eigs} that there is a line of magnetic field going straight through $r=0$ for the $m=1$ mode. 
While there is no unique direction normal to a point, this magnetic field line aligns with the $x$-axis; in this case, the $x$ direction is similar to the normal direction for a two dimensional current sheet.

We check for continuity of the magnetic field by plotting $B_x$ of the eigenmode along the $x$ axis (see the top panel of figure \ref{fig:diffbl}). The magnetic field is symmetric about the $x$-axis and continuous.
In the limit of small resistivity, the gradient of the magnetic perturbation should become discontinuous for a tearing-mode-like instability.
In the bottom panel of figure \ref{fig:diffbl}, we plot the magnetic perturbations near $r=0$, in the region denoted by the vertical lines. 
At the boundary layer, $B_x$ transitions from negative to positive curvature as the resistivity decreases. With the exception of $\eta=\num{1.3e-5}$ (where the curvature is close to zero) the boundary layer becomes increasingly sharp as resistivity decreases.  
This suggests that the derivative of the magnetic perturbation would become discontinuous in the limit of $\eta\rightarrow0$.
This trend toward a less smooth function suggests that we are seeing a tearing mode like instability of the Prendergast magnetic field. The growth rate of a tearing mode instability scales as $\gamma \propto \eta^{3/5}$, which is a steeper power law than we see in our simulations (see figure \ref{fig:nugamma}), indicating that while the tearing mode like behavior at $r=0$ contributes to the instability, it may not be the only factor. 

We note that the growth rates differ when the boundary conditions are changed. We find the boundary layer widths to be similar for the different boundary conditions.
While the concavity changes near $\eta \approx \num{1.3e-5}$ for potential boundary conditions,
it is always negative for perfectly conducting boundary conditions. As the boundary layer widths are similar, it appears that the  boundary layer behavior does not account for the difference in growth rates between the two boundary conditions.

In contrast to our simulations with $m=1$, we see in figure \ref{fig:m0_vis} that simulations with $m=0$ have different behavior at $r=0$. While we still have a guide field in the z-direction, the magnetic field is exactly zero in all components at $r=0$.
For a tearing mode instability to occur, the perturbation field must have a nonzero component normal to the guide field. The $m=0$ mode requires axisymmetry, meaning that a non-zero component of magnetic field at $r=0$ can only lie along the axis of symmetry, in this case the z-axis. The requirements of axisymmetry are incompatible with the requirements for the tearing mode instability, indicating that $m=0$ simulations do not have tearing mode like behavior at the origin.
Figure \ref{fig:nugamma} shows that the $m=0$ and $m=1$ modes have different behavior as a function of resistivity. At high resistivity, the $m=1$ modes show a steeper power law than the $m=0$ modes while at low resistivity the power laws have the same slope. This suggests the tearing mode like behavior affects the growth rate at high resistivity but not at low resistivity.

While we see instability in both the $m=0$ and $m=1$ cases, the tearing mode like behavior is only present in the $m=1$ simulations. This indicates that a tearing mode like mechanism can contribute to the instability of the Prendergast magnetic field, but is not a requirement for instability. 

\subsection{Extrapolating to astrophysical parameters}\label{irlstars}
Next we determine if this instability is relevant to physical stars. Realistic stellar diffusivities are too small to directly simulate. However, by fitting a trend between growth rate and diffusivity, we can extrapolate this trend down to the diffusivities seen in physical systems and determine the relevance of the instability. 
Figure \ref{fig:nugamma} shows the instability growth rates decrease as power-laws in the diffusivity, the slope of which changes as boundary conditions or Brunt-V{\"a}is{\"a}l{\"a} frequency change. We look at two different stable stratification strengths, noting that growth rate is roughly constant at high $N_0^2$ and thus the power-law scaling should also be constant over high $N_0^2$. The PC and POT lines are at $N_0^2=0$ and the $N^2$ line is at $N_0^2=1.69$ Power laws are given in terms of our non-dimensional variables, but in order to extrapolate to physical stars we dimensionalize the variables.
In each case we fit the data to the dimensionalized power law 
\begin{equation}
    \gamma = \alpha \left(\frac{\eta}{v_AR}\right)^{\beta} \frac{v_A}{R}
\end{equation}
where $R$ is the radius of the radiative region of the star, and $v_A$ is the Alfven velocity. Then the e-folding timescale $\tau$ of the instability is
\begin{equation}
    \tau = \frac{1}{\gamma} = \frac{1}{\alpha} \left(\frac{\eta}{v_AR}\right)^{-\beta} \frac{R}{v_A}
\end{equation}
 
Putting this in terms of the Lundquist number, $S=\frac{Rv_a}{\eta}$, we find that 
\begin{equation}
    \tau = \frac{1}{\alpha} S^{\beta} \frac{R}{v_a}
\end{equation}
From here, we can calculate the timescale of instability for a typical radiative zone of an RGB star, given a radius of ~0.66 $R_\odot$, mass of 1.6 $M_\odot$, magnetic field of $10^5 \, {\rm G}$ \citep[][]{astroseismology}, density of $10^5 \, {\rm g}/{\rm cm}^3$ and $\eta$ of 1 ${\rm cm}^2/{\rm s}$ \citep[][]{atlas}. 
Using $v_A = \frac{B^2}{4\pi\rho}$, we find $v_A$ of 90 ${\rm cm}/ {\rm s}$. 

We find different growth timescales for each of the fits reported in Fig.~\ref{fig:nugamma}; the longest timescale, corresponding to simulations with stable stratification is $\tau=$\num{1.2e6} years. 
We find the same growth timescale in all simulations with strong stable stratification.
As the cores of RGB stars are strongly stably stratified, we expect this to be the most physically relevant timescale.
All simulations with no stable stratification have shorter timescales. 
The lifetime of a RGB star is on the order of a few million years, so we expect the instability will grow to be relevant over the lifetime of the star. Because the Prendergast field is unstable over a fraction of the stellar evolutionary timescale, it is not a good model to use for studies of stable magnetic fields in stars.

\section{Conclusions}
Convection is thought to generate large scale, core magnetic fields, which persist in stars after they evolve off the main sequence. It’s thought that the remnants of these fields may take the form of the Prendergast magnetic field, a combination of poloidal and toroidal field components which are expected to stabilize each other. Previous analytic and numerical stability calculations  have suggested this magnetic field is stable. Our numerical calculations show a linear instability of this magnetic field on a timescale longer than previously studied. 

Section \ref{instability} presented the underlying instability. We confirmed that the instability was physical by showing a consistent growth rate across increasing spatial resolution and decreasing temporal resolution.  
In section \ref{diff} we discussed the effects of magnetic resistivity on the instability. We found that the growth rate decreased as resistivity decreased, indicating that our system exhibits a resistive instability. 

Section \ref{stablestrat} discussed the addition of stable stratification into the system, and found that the growth rate became constant at high strengths of stable stratification. We visualized the unstable eigenmodes and discussed the similarities in the magnetic eigenmodes despite a difference in velocity eigenmodes. 
Section \ref{mandBC} discussed the effect of changing azimuthal wavenumber and boundary conditions on the instability. We found that the $m=0$ and $m=1$ modes were unstable, while higher order modes were stable. We found the growth rates were larger when potential boundary conditions were used than when perfectly conducting boundary conditions were used.

Section \ref{r0} discussed the tearing mode instability as a possible mechanism for the Prendergast field instability.  
The eigenmodes of the tearing mode instability exhibit continuous magnetic field perturbations normal to the boundary layer and guide field, and a discontinuous normal derivative of the normal field across the boundary layer. 
We found that the eigenmodes of our $m=1$ simulations exhibit this behavior, while the eigenmodes of our $m=0$ simulations can not exhibit this behavior due to symmetry constraints. 
Our $m=1$ simulations with different boundary conditions have similar boundary layer widths at $r=0$, indicating that the boundary layer at $r=0$ is not responsible for the difference in growth rates seen in simualtions with different boundary conditions.
While the tearing mode instability may contribute to the instability in the Prendergast field for $m=1$, it is not a necessary component for instability of the Prendergast magnetic field. 

In section \ref{irlstars} we discussed implications of this instability for red giant stars. We calculated the e-folding timescale of the instability for stars, and found that it is much less than the evolution timescale, indicating that the Prendergast magnetic field is not a good model to study stable magnetic fields in stars.

Our calculations call into question the validity of past results which used the Prendergast field as a stable magnetic field.
The e-folding timescale of the instability is faster than the evolution timescale of a star, so the Prendergast field is a poor model for a stable magnetic field in stellar interiors. 
However, since the instability timescale is long compared to the Alfven timescale, it is likely that results obtained on a short timescale will see no effect of this instability. 

Future work should examine the nonlinear saturation of the instability. It is possible that in a nonlinear system this instability will saturate at low amplitude and thus have little effect on the system as a whole. Additionally, future work should aim to find a stable magnetic field that can be used to study magnetic waves in linear systems. Similar tests to the above work could be used to verify the stability of such a field.

\section*{Acknowledgements}

We thank Adam Jermyn for useful discussions on typical stellar parameters and Liam O'Connor, Ben Hyatt, Adrian Fraser, Kyle Augustson, and Whitney Powers for their valuable feedback and discussions.  

DL is supported in part by NASA HTMS grant 80NSSC20K1280.
EHA is funded as a CIERA Postdoctoral fellow and would like to thank CIERA and Northwestern University.
Computations were conducted with support from the NASA High End Computing (HEC) Program through the NASA Advanced Supercomputing (NAS) Division at Ames Research Center on Pleiades with allocation GID s2276.
This research was supported in part by the National Science Foundation under Grant No. NSF PHY-1748958.

\section*{Data Availability}

All scripts used to generate data and make the figures for this paper are available at https://github.com/ekaufman5/PrendergastScripts.



\bibliographystyle{mnras}
\bibliography{bibliography} 

\begin{thebibliography}{}
\makeatletter
\relax
\def\mn@urlcharsother{\let\do\@makeother \do\$\do\&\do\#\do\^\do\_\do\%\do\~}
\def\mn@doi{\begingroup\mn@urlcharsother \@ifnextchar [ {\mn@doi@}
  {\mn@doi@[]}}
\def\mn@doi@[#1]#2{\def\@tempa{#1}\ifx\@tempa\@empty \href
  {http://dx.doi.org/#2} {doi:#2}\else \href {http://dx.doi.org/#2} {#1}\fi
  \endgroup}
\def\mn@eprint#1#2{\mn@eprint@#1:#2::\@nil}
\def\mn@eprint@arXiv#1{\href {http://arxiv.org/abs/#1} {{\tt arXiv:#1}}}
\def\mn@eprint@dblp#1{\href {http://dblp.uni-trier.de/rec/bibtex/#1.xml}
  {dblp:#1}}
\def\mn@eprint@#1:#2:#3:#4\@nil{\def\@tempa {#1}\def\@tempb {#2}\def\@tempc
  {#3}\ifx \@tempc \@empty \let \@tempc \@tempb \let \@tempb \@tempa \fi \ifx
  \@tempb \@empty \def\@tempb {arXiv}\fi \@ifundefined
  {mn@eprint@\@tempb}{\@tempb:\@tempc}{\expandafter \expandafter \csname
  mn@eprint@\@tempb\endcsname \expandafter{\@tempc}}}

\bibitem[\protect\citeauthoryear{{Augustson}, {Brun}  \& {Toomre}}{{Augustson}
  et~al.}{2016}]{convection}
{Augustson} K.~C.,  {Brun} A.~S.,   {Toomre} J.,  2016, \mn@doi [\apj]
  {10.3847/0004-637X/829/2/92}, \href
  {https://ui.adsabs.harvard.edu/abs/2016ApJ...829...92A} {829, 92}

\bibitem[\protect\citeauthoryear{{Braithwaite} \& {Nordlund}}{{Braithwaite} \&
  {Nordlund}}{2006}]{Braithwaite+nordlund2005}
{Braithwaite} J.,  {Nordlund} {\r{A}}.,  2006, \mn@doi [\aap]
  {10.1051/0004-6361:20041980}, \href
  {https://ui.adsabs.harvard.edu/abs/2006A&A...450.1077B} {450, 1077}

\bibitem[\protect\citeauthoryear{{Bugnet} et~al.,}{{Bugnet}
  et~al.}{2021}]{bugnet2021}
{Bugnet} L.,  et~al., 2021, \mn@doi [\aap] {10.1051/0004-6361/202039159}, \href
  {https://ui.adsabs.harvard.edu/abs/2021A&A...650A..53B} {650, A53}

\bibitem[\protect\citeauthoryear{{Burns}, {Vasil}, {Oishi}, {Lecoanet}  \&
  {Brown}}{{Burns} et~al.}{2020}]{dedalus}
{Burns} K.~J.,  {Vasil} G.~M.,  {Oishi} J.~S.,  {Lecoanet} D.,   {Brown} B.~P.,
   2020, \mn@doi [Physical Review Research] {10.1103/PhysRevResearch.2.023068},
  \href {https://ui.adsabs.harvard.edu/abs/2020PhRvR...2b3068B} {2, 023068}

\bibitem[\protect\citeauthoryear{{Cantiello} \& {Braithwaite}}{{Cantiello} \&
  {Braithwaite}}{2011}]{heattransport}
{Cantiello} M.,  {Braithwaite} J.,  2011, \mn@doi [\aap]
  {10.1051/0004-6361/201117512}, \href
  {https://ui.adsabs.harvard.edu/abs/2011A&A...534A.140C} {534, A140}

\bibitem[\protect\citeauthoryear{{Charbonneau} \& {MacGregor}}{{Charbonneau} \&
  {MacGregor}}{2001}]{dynamofields1}
{Charbonneau} P.,  {MacGregor} K.~B.,  2001, \mn@doi [\apj] {10.1086/322417},
  \href {https://ui.adsabs.harvard.edu/abs/2001ApJ...559.1094C} {559, 1094}

\bibitem[\protect\citeauthoryear{{Cowling}}{{Cowling}}{1945}]{fossilfield1}
{Cowling} T.~G.,  1945, \mn@doi [\mnras] {10.1093/mnras/105.3.166}, \href
  {https://ui.adsabs.harvard.edu/abs/1945MNRAS.105..166C} {105, 166}

\bibitem[\protect\citeauthoryear{{Cowling}}{{Cowling}}{1960}]{cowling1960}
{Cowling} T.~G.,  1960, \mn@doi [\mnras] {10.1093/mnras/121.4.393}, \href
  {https://ui.adsabs.harvard.edu/abs/1960MNRAS.121..393C} {121, 393}

\bibitem[\protect\citeauthoryear{{Drake} \& {Lee}}{{Drake} \&
  {Lee}}{1977}]{drake1}
{Drake} J.~F.,  {Lee} Y.~C.,  1977, \mn@doi [Physics of Fluids]
  {10.1063/1.862017}, \href
  {https://ui.adsabs.harvard.edu/abs/1977PhFl...20.1341D} {20, 1341}

\bibitem[\protect\citeauthoryear{{Drake}, {Swisdak}, {Cattell}, {Shay},
  {Rogers}  \& {Zeiler}}{{Drake} et~al.}{2003}]{drake2}
{Drake} J.~F.,  {Swisdak} M.,  {Cattell} C.,  {Shay} M.~A.,  {Rogers} B.~N.,
  {Zeiler} A.,  2003, \mn@doi [Science] {10.1126/science.1080333}, \href
  {https://ui.adsabs.harvard.edu/abs/2003Sci...299..873D} {299, 873}

\bibitem[\protect\citeauthoryear{{Duez}, {Braithwaite}  \& {Mathis}}{{Duez}
  et~al.}{2010}]{Duez2010}
{Duez} V.,  {Braithwaite} J.,   {Mathis} S.,  2010, \mn@doi [\apjl]
  {10.1088/2041-8205/724/1/L34}, \href
  {https://ui.adsabs.harvard.edu/abs/2010ApJ...724L..34D} {724, L34}

\bibitem[\protect\citeauthoryear{{Fan}}{{Fan}}{2021}]{solardynamo}
{Fan} Y.,  2021, \mn@doi [Living Reviews in Solar Physics]
  {10.1007/s41116-021-00031-2}, \href
  {https://ui.adsabs.harvard.edu/abs/2021LRSP...18....5F} {18, 5}

\bibitem[\protect\citeauthoryear{{Fontana}, {Mininni}  \& {Dmitruk}}{{Fontana}
  et~al.}{2022}]{fontana}
{Fontana} M.,  {Mininni} P.~D.,   {Dmitruk} P.,  2022, arXiv e-prints, \href
  {https://ui.adsabs.harvard.edu/abs/2022arXiv220702394F} {p. arXiv:2207.02394}

\bibitem[\protect\citeauthoryear{{Fuller}, {Cantiello}, {Stello}, {Garcia}  \&
  {Bildsten}}{{Fuller} et~al.}{2015}]{astroseismology}
{Fuller} J.,  {Cantiello} M.,  {Stello} D.,  {Garcia} R.~A.,   {Bildsten} L.,
  2015, \mn@doi [Science] {10.1126/science.aac6933}, \href
  {https://ui.adsabs.harvard.edu/abs/2015Sci...350..423F} {350, 423}

\bibitem[\protect\citeauthoryear{{Furth}, {Killeen}  \& {Rosenbluth}}{{Furth}
  et~al.}{1963}]{tearing}
{Furth} H.~P.,  {Killeen} J.,   {Rosenbluth} M.~N.,  1963, \mn@doi [Physics of
  Fluids] {10.1063/1.1706761}, \href
  {https://ui.adsabs.harvard.edu/abs/1963PhFl....6..459F} {6, 459}

\bibitem[\protect\citeauthoryear{{Harrington} \& {Garaud}}{{Harrington} \&
  {Garaud}}{2019}]{chmicakmixing}
{Harrington} P.~Z.,  {Garaud} P.,  2019, \mn@doi [\apjl]
  {10.3847/2041-8213/aaf812}, \href
  {https://ui.adsabs.harvard.edu/abs/2019ApJ...870L...5H} {870, L5}

\bibitem[\protect\citeauthoryear{{Jermyn}, {Anders}, {Lecoanet}  \&
  {Cantiello}}{{Jermyn} et~al.}{2022}]{atlas}
{Jermyn} A.~S.,  {Anders} E.~H.,  {Lecoanet} D.,   {Cantiello} M.,  2022, arXiv
  e-prints, \href {https://ui.adsabs.harvard.edu/abs/2022arXiv220600011J} {p.
  arXiv:2206.00011}

\bibitem[\protect\citeauthoryear{{Landstreet}}{{Landstreet}}{1992}]{obs_bfield}
{Landstreet} J.~D.,  1992, \mn@doi [\aapr] {10.1007/BF00873569}, \href
  {https://ui.adsabs.harvard.edu/abs/1992A&ARv...4...35L} {4, 35}

\bibitem[\protect\citeauthoryear{{Lecoanet}, {Vasil}, {Burns}, {Brown}  \&
  {Oishi}}{{Lecoanet} et~al.}{2018}]{spheres_b}
{Lecoanet} D.,  {Vasil} G.~M.,  {Burns} K.~J.,  {Brown} B.~P.,   {Oishi} J.~S.,
   2018, arXiv e-prints, \href
  {https://ui.adsabs.harvard.edu/abs/2018arXiv180409283L} {p. arXiv:1804.09283}

\bibitem[\protect\citeauthoryear{Li, Jaroszynski, Pearse, Orf  \& Clyne}{Li
  et~al.}{2019}]{VAPOR}
Li S.,  Jaroszynski S.,  Pearse S.,  Orf L.,   Clyne J.,  2019, \mn@doi
  [Atmosphere] {10.3390/atmos10090488}, 10

\bibitem[\protect\citeauthoryear{{Loi}}{{Loi}}{2020}]{Loi}
{Loi} S.~T.,  2020, \mn@doi [\mnras] {10.1093/mnras/staa581}, \href
  {https://ui.adsabs.harvard.edu/abs/2020MNRAS.493.5726L} {493, 5726}

\bibitem[\protect\citeauthoryear{{Loi} \& {Papaloizou}}{{Loi} \&
  {Papaloizou}}{2017}]{loi2017}
{Loi} S.~T.,  {Papaloizou} J. C.~B.,  2017, \mn@doi [\mnras]
  {10.1093/mnras/stx281}, \href
  {https://ui.adsabs.harvard.edu/abs/2017MNRAS.467.3212L} {467, 3212}

\bibitem[\protect\citeauthoryear{{Loi} \& {Papaloizou}}{{Loi} \&
  {Papaloizou}}{2018}]{loi2018}
{Loi} S.~T.,  {Papaloizou} J. C.~B.,  2018, \mn@doi [\mnras]
  {10.1093/mnras/sty917}, \href
  {https://ui.adsabs.harvard.edu/abs/2018MNRAS.477.5338L} {477, 5338}

\bibitem[\protect\citeauthoryear{{Markey} \& {Tayler}}{{Markey} \&
  {Tayler}}{1973}]{poloidalstability}
{Markey} P.,  {Tayler} R.~J.,  1973, \mn@doi [\mnras] {10.1093/mnras/163.1.77},
  \href {https://ui.adsabs.harvard.edu/abs/1973MNRAS.163...77M} {163, 77}

\bibitem[\protect\citeauthoryear{Mitchell, Braithwaite, Langer, Reisenegger  \&
  Spruit}{Mitchell et~al.}{2013}]{mitchell2014}
Mitchell J.~P.,  Braithwaite J.,  Langer N.,  Reisenegger A.,   Spruit H.,
  2013, \mn@doi [Proceedings of the International Astronomical Union]
  {10.1017/S1743921314002701}, 9, 441–444

\bibitem[\protect\citeauthoryear{Neiner, Mathis, Alecian, Emeriau  \&
  Grunhut}{Neiner et~al.}{2014}]{massive_no_dynamo}
Neiner C.,  Mathis S.,  Alecian E.,  Emeriau C.,   Grunhut J.,  2014, \mn@doi
  [Proceedings of the International Astronomical Union]
  {10.1017/S1743921315004524}, 10, 61–66

\bibitem[\protect\citeauthoryear{{Prat}, {Mathis}, {Buysschaert}, {Van Beeck},
  {Bowman}, {Aerts}  \& {Neiner}}{{Prat} et~al.}{2019}]{prat2019}
{Prat} V.,  {Mathis} S.,  {Buysschaert} B.,  {Van Beeck} J.,  {Bowman} D.~M.,
  {Aerts} C.,   {Neiner} C.,  2019, \mn@doi [\aap]
  {10.1051/0004-6361/201935462}, \href
  {https://ui.adsabs.harvard.edu/abs/2019A&A...627A..64P} {627, A64}

\bibitem[\protect\citeauthoryear{{Prendergast}}{{Prendergast}}{1956}]{Prendergast1956}
{Prendergast} K.~H.,  1956, \mn@doi [\apj] {10.1086/146186}, \href
  {https://ui.adsabs.harvard.edu/abs/1956ApJ...123..498P} {123, 498}

\bibitem[\protect\citeauthoryear{{Prendergast}}{{Prendergast}}{1958}]{Prendergast2}
{Prendergast} K.~H.,  1958, \mn@doi [\apj] {10.1086/146549}, \href
  {https://ui.adsabs.harvard.edu/abs/1958ApJ...128..361P} {128, 361}

\bibitem[\protect\citeauthoryear{{Stello}, {Cantiello}, {Fuller}, {Garcia}  \&
  {Huber}}{{Stello} et~al.}{2016}]{stellocantiello}
{Stello} D.,  {Cantiello} M.,  {Fuller} J.,  {Garcia} R.~A.,   {Huber} D.,
  2016, \mn@doi [\pasa] {10.1017/pasa.2016.9}, \href
  {https://ui.adsabs.harvard.edu/abs/2016PASA...33...11S} {33, e011}

\bibitem[\protect\citeauthoryear{{Tayler}}{{Tayler}}{1973}]{toroidal_stablility}
{Tayler} R.~J.,  1973, \mn@doi [\mnras] {10.1093/mnras/161.4.365}, \href
  {https://ui.adsabs.harvard.edu/abs/1973MNRAS.161..365T} {161, 365}

\bibitem[\protect\citeauthoryear{{Thelen} \& {Cattaneo}}{{Thelen} \&
  {Cattaneo}}{2000}]{Thelen}
{Thelen} J.~C.,  {Cattaneo} F.,  2000, \mn@doi [\mnras]
  {10.1046/j.1365-8711.2000.03620.x}, \href
  {https://ui.adsabs.harvard.edu/abs/2000MNRAS.315L..13T} {315, L13}

\bibitem[\protect\citeauthoryear{{Varela}, {Brun}, {Dubrulle}  \&
  {Nore}}{{Varela} et~al.}{2017}]{varela}
{Varela} J.,  {Brun} S.,  {Dubrulle} B.,   {Nore} C.,  2017, \mn@doi [Physics
  of Plasmas] {10.1063/1.4983313}, \href
  {https://ui.adsabs.harvard.edu/abs/2017PhPl...24e3518V} {24, 053518}

\bibitem[\protect\citeauthoryear{{Vasil}, {Lecoanet}, {Burns}, {Oishi}  \&
  {Brown}}{{Vasil} et~al.}{2018}]{spheres_a}
{Vasil} G.,  {Lecoanet} D.,  {Burns} K.,  {Oishi} J.,   {Brown} B.,  2018,
  arXiv e-prints, \href {https://ui.adsabs.harvard.edu/abs/2018arXiv180410320V}
  {p. arXiv:1804.10320}

\bibitem[\protect\citeauthoryear{Wang \& Ruuth}{Wang \& Ruuth}{2008}]{SBDF2}
Wang D.,  Ruuth S.~J.,  2008, Journal of Computational Mathematics, 26, 838

\bibitem[\protect\citeauthoryear{{Weber} \& {Davis}}{{Weber} \&
  {Davis}}{1967}]{stellarwinds}
{Weber} E.~J.,  {Davis} Leverett J.,  1967, \mn@doi [\apj] {10.1086/149138},
  \href {https://ui.adsabs.harvard.edu/abs/1967ApJ...148..217W} {148, 217}

\makeatother
\end{thebibliography}





\bsp	
\label{lastpage}
\end{document}